# Coupling between a dark and a bright eigenmode in a terahertz metamaterial

**Ranjan Singh,**[1]  **Carsten Rockstuhl,**[2] **Falk Lederer,**[2] **and Weili Zhang**[1,*]

[1] *School of Electrical and Computer Engineering, Oklahoma State University, Stillwater, Oklahoma 74078, USA*
[2] *Institute of Condensed Matter Theory and Solid State Optics, Friedrich-Schiller-Universität Jena, Jena 07743, Germany*

## Abstract

Terahertz time domain spectroscopy and rigorous simulations are used to probe the coupling between a dark and a bright plasmonic eigenmode in a metamaterial with broken symmetry. The metamaterial consists of two closely spaced split ring resonators that have their gaps in non-identical positions within the ring. For normal incidence and a fixed polarization both lowest order eigenmodes of the split ring resonators can be excited; although one of them has to be regarded as dark since coupling is prohibited because of symmetry constraints. Emphasis in this work is put on a systematic evaluation of the coupling effects depending on a spectral tuning of both resonances.

Metamaterials are a recently introduced novel class of artificial matter that is composed of, usually, periodically arranged unit cells [1]. Their purpose is to control the properties of light propagation at will. This is accomplished by defining an appropriately tailored geometry for the unit cell. If the unit cells are sufficiently small, it is possible to homogenize the structure and attribute effective material parameters [2]. To induce an effective electric and/or magnetic polarization that strongly deviates from that of a simple spatial average of the intrinsic material properties, resonances are usually evoked. Prominent examples for such unit cells are cut wire pairs [3] or split ring resonators (SRR) [4]. At optical frequencies the resonances rely on the excitation of localized plasmon polaritons in the metallic structures forming the entities in the unit cell [5]. At lowered wavelengths the same resonances persist, though then they should be rather understood on the base of an antenna effect. Once the fundamental mechanisms were revealed, sudden interest sparked on how the concepts of coupling between plasmonic eigenmodes can be exploited to sculpture the resonances of metamaterials in a more appropriate manner. Prominent examples are the application of the plasmon hybridization method [6] to explain the occurrence of magnetic atoms [7] or even, more recently, magnetic molecule [8]. Fabricated elements for this purpose evoke an appropriate coupling between laterally stacked SRRs [9]. Such investigations are usually driven by the desire to enlarge the bandwidth or to sharpen the resonances. As a rule of thumb one is inclined to believe that the sharper the resonance the stronger is the induced dispersion in the effective material parameters. Such resonance sharpening is extremely beneficial for obtaining low loss metamaterials. As the imaginary part of the material parameter decays off resonance faster than the real part, such sharp resonances would permit to choose an operating frequency for the metamaterial far away from the resonance position.

Potentially the most appealing approach to observe sharp resonances in coupled objects makes use of the excitation of a polaritonic resonance that is formed between a bright and a dark eigenmode [10]. For the purpose of coupling, the symmetry of the system has to be broken in order to allow the excitation of the dark state. Otherwise, it would have been forbidden. First investigations on such kind of systems were recently reported. They were designed to operate either at optical [10] or at radio frequencies [11]. Also suggestions on how to use SRRs to observe such kind of interaction phenomena were put forward [12]. Due to the apparent similarity to quantum interference in an atomic system comprising two indistinguishable paths, these investigations were regarded as a plasmonic analogy to electromagnetic induced transparency [13]. Whereas in atomic systems the spectral detuning of both resonances is understood as the principal feature that ultimately tailors the resonance, thus far no investigation was reported on how such detuning affects the coupled states in their plasmonic counterparts. It has to be made furthermore clear that the immediate application of the analogy between atomic and plasmonic systems to observe EIT like phenomena is inappropriate, as long as the relative line widths of the resonances that can be excited in fabricatable plasmonic devices are not chosen properly. A true analogy to EIT requires the line width of both involved resonances to be different in magnitude. The bright eigenmode has to be broad to allow for a good absorption of light, whereas the dark mode has to have a narrow line width [14]. It remains open whether this condition can be met using SRRs.

To provide an answer to the open questions that have been formulated above we perform here a systematic investigation of the impact of the relative spectral position of the involved resonances on the polaritonic coupled state. We distinguish scenarios of weak coupling and strong coupling. For this purpose we focus on pairs of SRRs. The length of the wires forming the SRR allows to control the spectral position of the resonances in these systems; the spacing between the SRRs permits to tune the coupling strength in a controlled manner. The spectral response of appropriately and systematically designed samples is experimentally measured using a devoted terahertz time domain spectroscopy (THz-TDS) setup **[15,16]**. Complementary rigorous numerical tools are employed to elucidate the effects theoretically. We have to stress that the main purpose of this investigation is to detail the phenomena occurring in the experimentally accessible spectral response from such structures. In consequence, no particular emphasis is put on the effective material parameters that could potentially be attributed to the medium.

The sample to be characterized is placed at the 3.5 mm diameter waist of the free space terahertz beam **[16]**. Two sets of samples are fabricated using conventional photolithography on an n-type 640 µm thick silicon wafer with 12 Ω cm resistivity. Their general layout along with the definition of all geometrical quantities is shown in Figure 1. The first set of samples, MM1-MM3 consists of touching SRRs. The section of the arm that both SRRs do share was merged into a single wire as evident from Figure 1a, which represents the array of MM1. The second set of samples, MM4-MM6 comprises pairs of SRRs separated by a distance of $s = 3$ µm. Figures 1b and 1c show the schematic and the detailed definition of all geometrical parameters for MM4-MM6, respectively. The peculiarity of all the samples is the different position of the gap within the ring in both SRRs. Each set consists of samples in which the first SRR dimensions are fixed. The exact parameters of its geometry as revealed in Figure 1c are $g = 2$ µm, $w = 36$ µm, $l' = 36$ µm, $t = 6$ µm, and $h = 200$ nm. The SRRs consist of vacuum deposited aluminum. Within each set of samples all the geometrical parameters of the second SRR were kept identical except its arm length. The length parameter $l$ was subject to variation and was set to be 36 µm, 51 µm, and 21 µm, respectively. The periodicity of the unit cells for all 6 structures is $\Lambda_x = 50$ µm by $\Lambda_y = 100$ µm. Each of the MM samples has a 10 mm x 10 mm clear aperture and the terahertz wave illuminates the structure at normal incidence. The polarization of the electric field is chosen to be parallel to the gap of the first SRR as indicated in Figure 1c. This configuration allows to excite in the spectral domain of interest only an eigenmode in the first SRR [17]. This eigenmode is bright and represents the lowest order odd eigenmode. By contrast, the lowest eigenmode that can be excited for the second SRR appears at higher frequencies outside the spectral domain of interest at 1.33 THz. It is the lowest order even eigenmode. Nonetheless, this lowest order eigenmode is actually the second order one of the structure as the first order eigenmode appears to be dark and cannot be excited with the chosen polarization. The presence of the first SRR, however, breaks the symmetry and the mode becomes excitable. In the present investigation the arm length $l$ that is subject to variation from sample to sample constitutes the parameter that permits to tune the spectral position of the dark relative to the bright mode. The separation $s$ is the parameter that allows controlling the coupling strength between these two eigenmodes.

Figure 2a shows the transmission spectra of the sample MM1. The unit cell of this sample contains two identical touching SRRs. The second SRR is rotated by 90 degrees with respect to the first one. The transmission spectrum is obtained by normalizing the measured transmission to the reference transmission of a blank n-type silicon wafer identical to the sample substrate. It is hence defined as $|E_s(\omega)/E_r(\omega)|$, where $E_s(\omega)$ and $E_r(\omega)$ are Fourier transformed time traces of the transmitted electric fields of the signal and the reference pulses, respectively. Well pronounced resonances are difficult to resolve as the odd modes of the first SRR and the even modes of the second SRR are both excited with the chosen polarization. Consequently, the resulting spectrum consists of a series of closely spaced resonances that add up to such a weakly modulated spectrum, although the spectral positions of the resonances can be identified as weak dips. The only resonance that appears well pronounced occurs in the spectral domain of 0.4 to 0.6 THz. Figure 2b shows this spectral domain zoomed. From separate simulations it can be deduced that the resonance position coincides with the lowest order odd eigenmode that can be excited in the first SRR with the chosen polarization. However, we clearly observe a doublet rather than a single resonance, resulting in a transparency peak at 0.5 THz. From preliminary considerations we can deduce that the doublet occurs because of the strong coupling between the bright and the dark plasmonic eigenmode. For complementary purpose the figures show also results from a rigorous numerical simulation of the experimental situation. Simulation is based on the Fourier modal method that takes into account the exact geometrical parameters, the dispersive permittivity of Al and the measurement procedure [18]. We observe an excellent agreement in the entire spectral response that was simulated.

To elucidate this coupling in detail, Figure 3a shows the transmission spectrum of samples MM1-MM3, which is the case of the touching SRRs. The length $l$ of the side arm of the second SRR for one sample was chosen such that for the isolated SRRs the resonance frequencies of the bright and the dark mode are the same. For the two other samples the dark mode appears at lower (higher) frequencies for a longer (shorter) length of its arms $l$. A pronounced and very strong spectral splitting of the bright mode is only observed for the sample where the wire length $l$ of both SRRs is the same. As the length of the second SRR is increased to 51 µm the dark eigenmode is excited at 0.37 THz. When the length is reduced to 21 µm in MM3, the dark eigenmode is shifted to the higher frequency of 0.77 THz. The excitation of the dark resonance is extremely weakened as soon as the size of the second SRR deviates strongly from the size of the first SRR. For comparison, Figure 3b shows results of the simulated transmission for the samples MM1-MM3. Simulations are in perfect agreement with the experimental data of Figure 3a. All qualitative features are resolved and, to a certain extent, even quantitatively. The remaining discrepancies occur because of minor deviations in the dimensions of the fabricated sample geometry when compared to the design values that were considered in the simulations. From supportive simulation it can be seen that the spectra are sensitive against slight variations in the gap width $g$ and the separation $s$.

Figure 3c shows the measured transmission spectrum of the 3 µm separated SRRs, MM4-MM6. Although difficult to resolve, the dark mode is weakly excited at 0.36 THz in sample MM5 ($l > l'$) and at 0.81 THz in sample MM6 ($l < l'$). The spectral separation of the two entities forming the doublet in sample MM4 ($l = l'$) is not as strong as for the touching SRRs. The numerically calculated spectrum is shown in Figure 3d. It is similarly in good agreement as the measurements.

Prior to discussing the results in detail, we show in Figure 4a the color-coded simulated transmitted energy for the scenario where the two SRRs are touching. In the simulation the length parameter $l$ is gradually increased from 15 to 50 µm, all the other geometrical parameters are set to be identical to their experimental counterparts. The extracted resonance positions are shown additionally (blue curve). The spectral position of the dark mode in the uncoupled scenario is likewise shown (green curve). The spectral positions were extracted from separate simulations of the same geometrical situation in the absence of the first SRR and a polarization of the incident electric field that was set to be parallel to the gap of the second SRR. This rotation of the incident polarization basically switches the eigenmode from being dark to bright. In passing we note that the spectral position of the bright mode remains constant and is evidently not affected by the variation of the wire length *l* of the second SRR. Figure 4b shows the same results for the samples where the two SRRs are separated by 3 µm.

From the experimental and the theoretical results we clearly observe a strong coupling between the bright and the dark plasmonic eigenmodes. Their interaction becomes possible because at a relative rotation by 90° the orientation of both SRRs breaks the overall symmetry of the coupled structure. This broken symmetry renders the excitation of the otherwise forbidden dark mode possible. However, due to their coupled nature the eigenmodes themselves are not excitable but rather a polaritonic state, which can be clearly seen from the simulated dispersion relation. For large spectral detuning of both resonances the excitation of the dark mode is rather weak; though traces in transmission remain evident. The closer the resonance frequencies get the stronger is the excitation of the dark mode. Similarly, the excited eigenmodes deviate from the spectral position of the eigenmodes for the isolated SRRs. Anticrossing causes a strong spectral separation of the bright and the dark mode from their spectral position in the unperturbed situation. We observe the excitation of a polaritonic state. A Rabi splitting of the energetic levels is experimentally observed and numerically verified for the case of the two touching SRRs. Such strong splitting renders the observed spectra to be similar to an Autler-Townes-like doublet **[19]**. On the contrary, if the coupling between both resonances is significantly reduced (MM4-MM6 where the SRRs are separated by $s = 3$ µm) we do not observe sharp spectral features in the transmission spectra, neither in the experiments nor in the simulations. Although qualitatively, all features remain the same when compared to the case of two touching SRRs, the interaction of both SRRs is fairly weak. The weak coupling seems to be insufficient to induce sudden spectral changes, as one would expect in the case of a truly electromagnetically induced transparence analogy. The reduced coupling is caused by the reduced spatial overlap of the two eigenmodes.

Because the eigenmodes are locally confined to the SRR, the probability to excite the dark mode is reduced in the broken symmetry configuration for a large separation *s* of both SRRs. From the experimental and the theoretical data we can safely explain all spectral phenomena on the base of the plasmon hybridization model. Because of the broken symmetry the dark mode is bright. We assume that no significant deviation of its quality factor is encountered as compared to the bright mode. If both modes couple, a spectral splitting occurs. It leads to the formation of a symmetric and an anti symmetric mode. And finally the magnitude of the observed spectral splitting depends on the coupling strength between both eigenmodes. The absence of phenomena that bear similarities to electromagnetically induced transparency can be explained following the discussion provided in the introduction. For their observation the line width of the resonances excited in the system have to have different orders of magnitude. Great care has to be devoted to this question, though it can be reached using an appropriate design for the plasmonic system **[20]**. In the present scenario, however, of two similar SRRs with the specific geometrical features, the line width of the resonances is about the same and the anticipated effects associated with EIT are not observed.

To summarize, our most important achievement in this work is to having elucidated the coupling between a dark and a bright plasmonic eigenmode in unit cells of metamaterials with broken symmetry. By increasing or decreasing the arm length of the second SRR the relative spectral position of both resonances can be controlled. It allows to investigate the dispersion relation of the polaritonic state that is formed between the dark and the bright mode. It was observed that for a noticeable excitation of the polaritonic state the spectral positions of both states have to be sufficiently close. A significant coupling strength between both states leads then to a strong spectral splitting which in turn results in a transparency peak. The interaction between a dark and a bright mode in such broken symmetry unit cells adds a complementary aspect to the great variety offered by nanophotonics to tailor spectral resonances at will. Particularly in the field of MM it might lead to the development of broad band unit cells, low loss metamaterials or spectrally strong dispersive unit cells having a large effective group index.

The authors acknowledge invaluable experimental assistance from Xinchao Lu and Jianqiang Gu, and fruitful discussions with Yongyao Chen and Zhen Tian. This work was partially supported by the US National Science Foundation and the German Federal Ministry of Education and Research (Metamat). Some computations utilized the IBM p690 cluster JUMP of the Forschungszentrum in Jülich, Germany.

———————————————

\* Corresponding author.

weili.zhang@okstate.edu

**Figure Captions**

FIG. 1 (color online). (a) Principal sketch of samples MM1 – MM3 where $s = 0$ µm and (b), principal sketch of samples MM4 –MM6 where $s = 3$ µm. (c) Detailed definition of the geometrical parameters at the example of the unit cells for MM4-MM6. They are chosen to be $t = 6$ µm, $g = 2$ µm, $l' = w = 36$ µm, $h = 200$ nm, $s = 3$ µm and $l$ is subject to variations. The periodicity of the unit cells in all samples, MM1-MM6 is $\Lambda_x = 50$ µm by $\Lambda_y = 100$ µm.

FIG. 2 (color online). Measured and simulated amplitude transmission spectra of MM1 array which consists of touching SRRs with identical arm lengths. The incident E field is polarized parallel to the gap of the first SRR in the unit cell.

FIG. 3 (color online). (a) Measured and (b) simulated amplitude transmission spectra of samples MM1-MM3. (c) Measured and (d) simulated transmission spectra of MM4-MM6.

FIG. 4 (color online). Simulated transmission of (a) touching SRRs and (b) 3 µm separated SRRs with the arm length $l$ of the second SRR changes from 15 to 50 µm. The green curve is the resonance position of the second SRR with the E field polarized parallel to its gap as extracted from separate simulations. The blue curve is the resonance position of the excited eigenmodes extracted from the spectra.

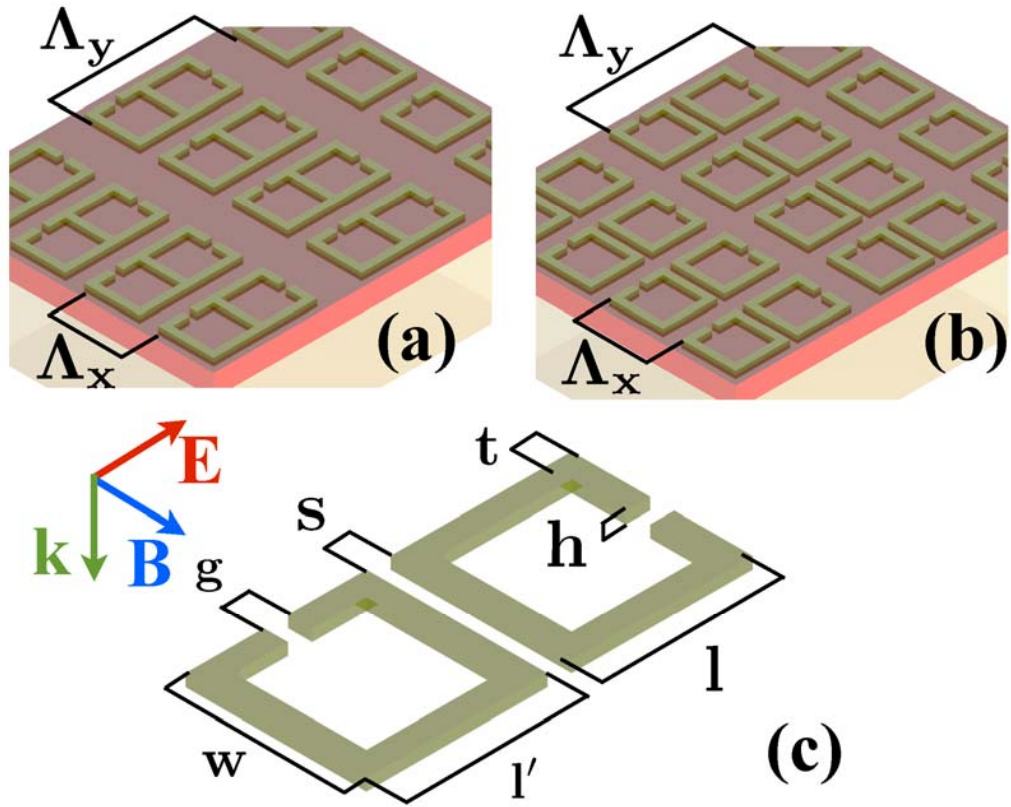

**Figure 1.**
**Singh** *et al.*

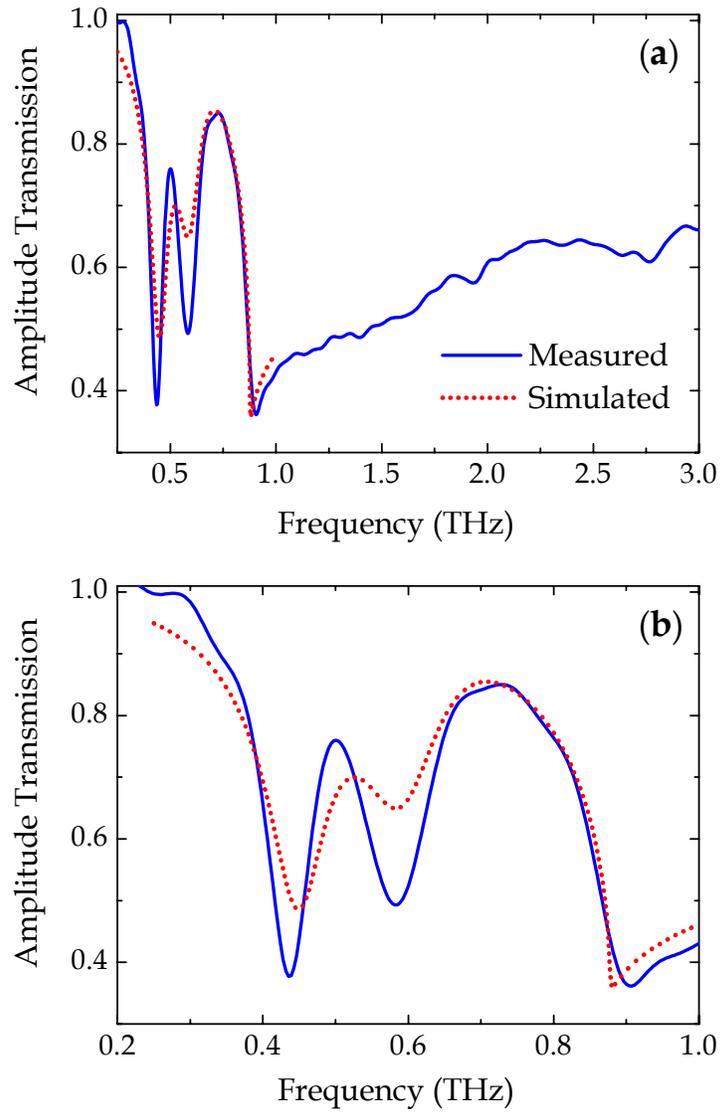

**Figure 2.**
**Singh** *et al.*

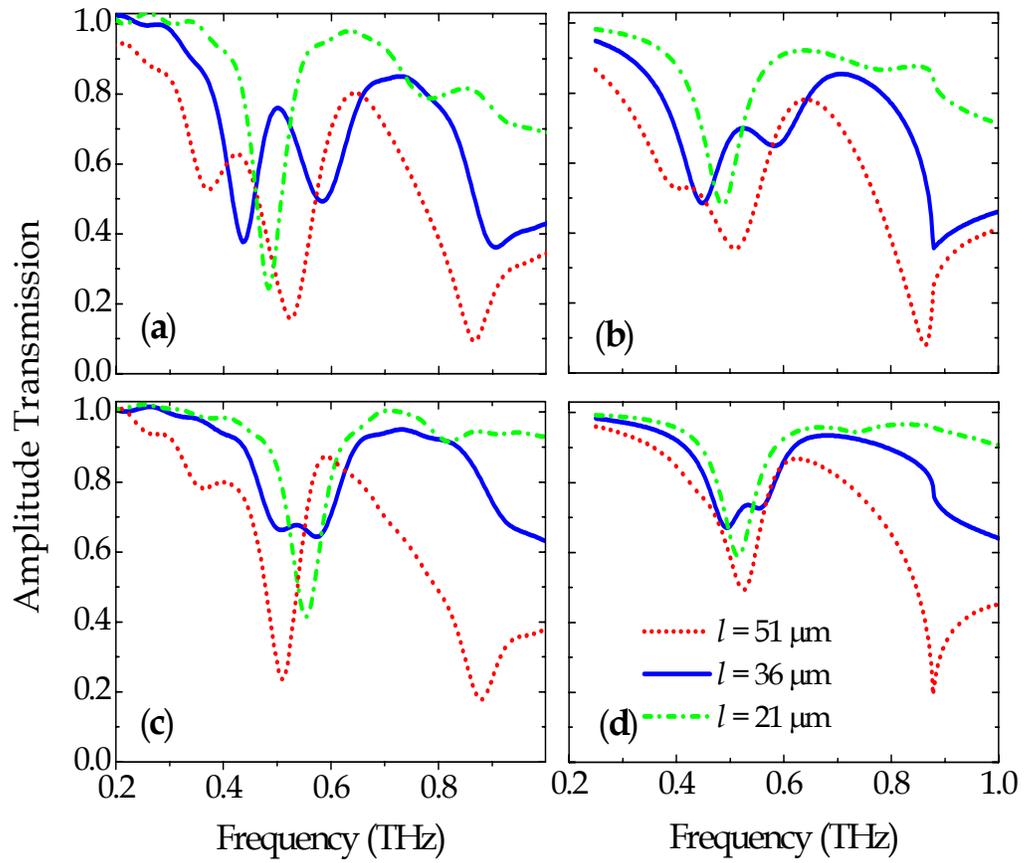

**Figure 3.
Singh *et al.***

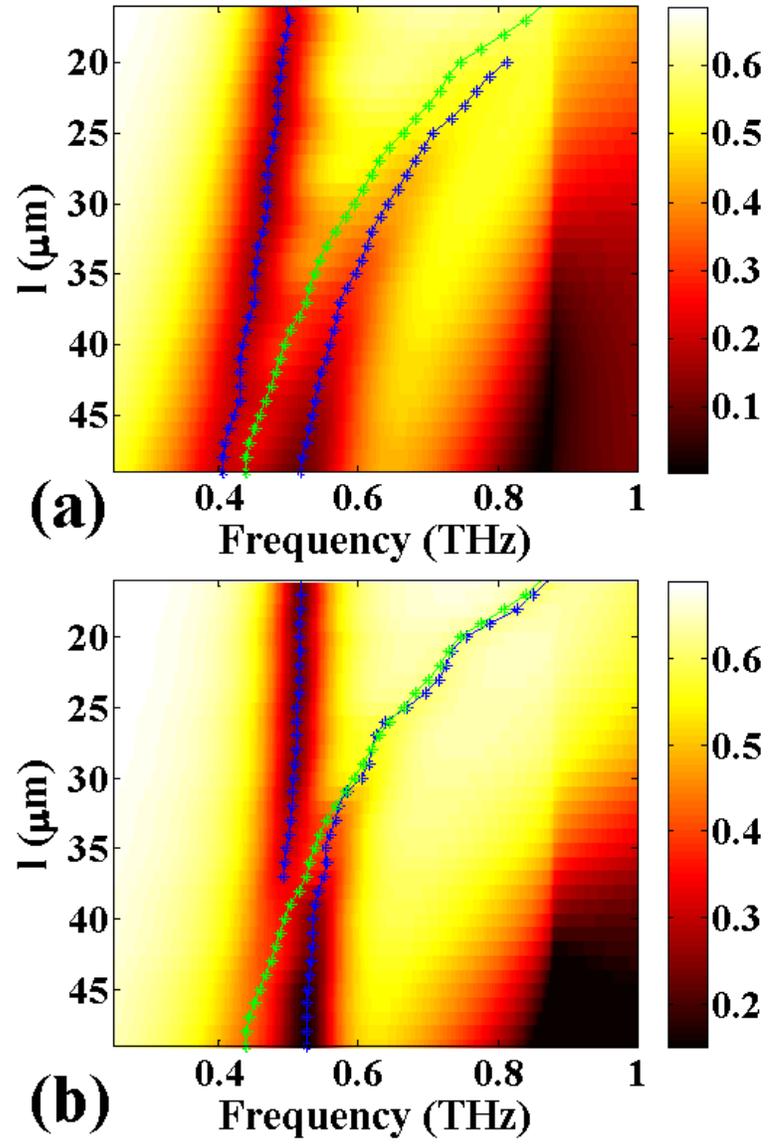

**Figure 4.**
**Singh** *et al.*

17